**Some remarks on point split commutators**

Dan Solomon
Rauland-Borg Corporation
3450 W. Oakton
Skokie, IL 60077  USA

Email: **dan.solomon@rauland.com**

(Nov 23,  2008)

**Abstract**

Point splitting has been suggested as a way to deal with anomalous commutators in quantum field theory.  It has been pointed out by D.G. Boulware[4] that in order to obtain a mathematically consistent theory the Hamiltonian operator must be point split also.  We will examine the effect of point splitting the Hamiltonian for a free fermion field in 1-1D space-time.  It will be shown that when the Hamiltonian operator is point split then quantum states will exist with less energy than the normal vacuum state.  This requires the vacuum state to be redefined.

**1. Introduction**

It is well know that quantum field theory contains anomalies [1][2]. An anomaly occurs when different ways of calculating a given quantity produce different results. An example of such an anomaly is the Schwinger term which is defined by the equal-time commutation relationship $\left\langle 0 \left| \left[ \hat{\rho}(\vec{x}), \hat{\vec{J}}(\vec{y}) \right] \right| 0 \right\rangle$ where $|0\rangle$ is the vacuum state and where the charge operator $\hat{\rho}(\vec{x})$ is defined by,

$$\hat{\rho}(\vec{x}) = \hat{\psi}^{\dagger}(\vec{x})\hat{\psi}(\vec{x}) \tag{1}$$

and the current operator is defined by,

$$\hat{\vec{J}}(\vec{x}) = \hat{\psi}^{\dagger}(\vec{x})\vec{\alpha}\hat{\psi}(\vec{x}) \tag{2}$$

where $\hat{\psi}$ is the field operator. The field operators obey the equal time anti-commutation relationship,

$$\left\{ \hat{\psi}_{\alpha}^{\dagger}(\vec{x}), \hat{\psi}_{\beta}(\vec{y}) \right\} = \delta_{\alpha\beta}\delta^{(3)}(\vec{x} - \vec{y}) \tag{3}$$

When the above relationships are used it can be shown that $\left\langle 0 \left| \left[ \hat{\rho}(\vec{x}), \hat{\vec{J}}(\vec{y}) \right] \right| 0 \right\rangle = 0$ However, it has been shown by J. Schwinger [3], and as discussed in [2], that if this relationship is true then there exist states with less energy than the vacuum state. Therefore, in order to preserve the fact that the vacuum state is a lower bound to the energy the Schwinger term must not be zero. One possible way to resolve this problem is to modify the definition of the current operator by using point splitting [3][4]. In point splitting the current operator is re-defined as,

$$\hat{\vec{J}}(\vec{x}; \vec{\varepsilon}) = \hat{\psi}^{\dagger}(\vec{x} + \vec{\varepsilon})\vec{\alpha}\hat{\psi}(\vec{x}) \tag{4}$$

where $\vec{\varepsilon}$ approaches zero in a symmetrical way. When this is done is found that it can be readily shown that $\left\langle 0 \left| \left[ \hat{\rho}(\vec{x}), \hat{\vec{J}}(\vec{y}; \vec{\varepsilon}) \right] \right| 0 \right\rangle \neq 0$. Therefore using point splitting to redefine the current operator per equation (4) resolves the anomaly involving the Schwinger term. However D.G. Boulware [4] has pointed out that, for a self consistent theory, point splitting must be introduced into other quantities. For example, in quantum field theory the continuity equation is given by,

$$\left[ \hat{H}, \hat{\rho}\left( \vec{x} \right) \right] = i\nabla \cdot \hat{\vec{J}}\left( \vec{x} \right) \tag{5}$$

Now if we use point splitting then $\hat{\vec{J}}\left( \vec{x} \right)$ is replaced by $\hat{\vec{J}}\left( \vec{x}; \vec{\varepsilon} \right)$ so that the right side of the equation is dependent on $\vec{\varepsilon}$. Therefore the left hand side must be dependent on $\vec{\varepsilon}$ also. This means that either $\hat{H}$ or $\hat{\rho}\left( \vec{x} \right)$ (or both) must be dependent on $\vec{\varepsilon}$. It is shown in [4] that the point splitting must be associated with the Hamiltonian and that the charge operator is not affected. Therefore the Hamiltonian will be written as $\hat{H}_{\vec{\varepsilon}}$ and equation (5) becomes,

$$\left[ \hat{H}_{\vec{\varepsilon}}, \hat{\rho}\left( \vec{x} \right) \right] = i\nabla \cdot \hat{\vec{J}}\left( \vec{x}; \vec{\varepsilon} \right) \tag{6}$$

In order for the theory to be internally consistent we specify $\hat{H}_{\vec{\varepsilon}}$ and then define $\hat{\vec{J}}\left( \vec{x}; \vec{\varepsilon} \right)$ by the above relationship.

In this paper we will examine some consequences of redefining the Hamiltonian using point splitting. To simplify the discussion we will consider a free fermion field in 1-1 dimension space-time. The discussion will proceed as follows. In Section 2 the basic elements of quantum field theory will be introduced. Then in Section 3 we will show how these elements lead to an anomaly. In Section 4 it will be shown that this anomaly can be resolved by point splitting the current operator. In Section 5 we will follow Boulware's suggestion and apply point splitting to the Hamiltonian. In Section 6 the effect of point splitting on the vacuum state will be examined. It will be shown that when the Hamiltonian is point split there will exist quantum states with less energy than the vacuum state. In order to resolve this problem the vacuum state must be redefined.

## 2. Basic elements of quantum field theory.

In this section we will write down the basic elements of a free quantized fermion field in 1-1dimension space-time where $z$ is the space dimension. The Hamiltonian operator is defined by,

$$\hat{H}_0 = \int \hat{\psi}^\dagger \left( z \right) H_0 \hat{\psi}\left( z \right) dz + \xi_R \tag{7}$$

where $\xi_R$ is a renormalization constant so that the energy of the vacuum state will be zero and,

$$H_0 = \left( -i\sigma_x \frac{\partial}{\partial z} + m\sigma_z \right) \tag{8}$$

where $\sigma_x$ and $\sigma_z$ are the usual Pauli matrices and $m$ is the mass.

The field operators obeys the anti-commutation relationship,

$$\hat{\psi}_\alpha^\dagger(z)\hat{\psi}_\beta(z') + \hat{\psi}_\beta(z')\hat{\psi}_\alpha^\dagger(z) = \delta_{\alpha\beta}\delta(z - z') \tag{9}$$

Using the above we can obtain the following useful relationship,

$$\left[ \hat{\psi}^\dagger(z)\hat{\psi}(z), \hat{\psi}^\dagger(z')\sigma_x\hat{\psi}(z'') \right] = \hat{\psi}^\dagger(z)\sigma_x\hat{\psi}(z'')\delta(z - z') - \hat{\psi}^\dagger(z')\sigma_x\hat{\psi}(z)\delta(z - z'')$$
$$\tag{10}$$

The field operators are defined by,

$$\hat{\psi}(z) = \int_{-\infty}^{+\infty} \left( \hat{b}_p\phi_{1,p}(z) + \hat{d}_p^\dagger\phi_{-1,p}(z) \right) dp; \quad \hat{\psi}^\dagger(z) = \int_{-\infty}^{+\infty} \left( \hat{b}_p^\dagger\phi_{1,p}^\dagger(z) + \hat{d}_p\phi_{-1,p}^\dagger(z) \right) dp \tag{11}$$

where the $\hat{b}_p$ ($\hat{b}_p^\dagger$) are the destruction(creation) operators for an electron associated with the state $\phi_{1,p}(z)$ and the $\hat{d}_p$ ($\hat{d}_p^\dagger$) are the destruction(creation) operators for a positron associated with the state $\phi_{-1,p}(z)$. They satisfy the anti-commutation relationships,

$$\left\{ \hat{d}_p, \hat{d}_{p'}^\dagger \right\} = \delta(p - p'); \left\{ \hat{b}_p, \hat{b}_{p'}^\dagger \right\} = \delta(p - p') \tag{12}$$

where all other anti-commutators are zero. The functions $\phi_{\lambda,p}(z)$ are solutions of,

$$H_0\phi_{\lambda,p}(z) = \lambda E_p\phi_{\lambda,p}(z) \tag{13}$$

where $\lambda = \pm 1$ is the sign of the energy and,

$$\phi_{\lambda,p}(z) = u_{\lambda,p}e^{ipz} \tag{14}$$

where $p$ is the momentum and where,

$$E_p = +\sqrt{p^2 + m^2} \; ; \; u_{\lambda,p} = N_{\lambda,p}\left( \begin{array}{c} 1 \\ p/(\lambda E_p + m) \end{array} \right); \; N_{\lambda,p} = \sqrt{\frac{\lambda E_p + m}{2\lambda E_p}} \tag{15}$$

The $\phi_{\lambda,p}(z)$ form an orthonormal basis set and satisfy,

$$\int \phi_{\lambda,p}^\dagger(z)\phi_{\lambda',p'}(z)dz = \delta_{\lambda\lambda'}\delta(p - p') \tag{16}$$

The vacuum state $|0\rangle$ is defined by,

$$\hat{d}_p \left|0\right\rangle = \hat{b}_p \left|0\right\rangle = 0 \text{ and } \left\langle 0\right|\hat{d}_p^\dagger = \left\langle 0\right|\hat{b}_p^\dagger = 0 \text{ for all } p \tag{17}$$

Use the above relationships in (7) to obtain,

$$\hat{H}_0 = \int_{-\infty}^{+\infty} E_p \left(\hat{b}_p^\dagger \hat{b}_p - \hat{d}_p \hat{d}_p^\dagger\right) dp + \xi_R \tag{18}$$

Next use (12) to obtain,

$$\hat{H}_0 = \int_{-\infty}^{+\infty} E_p \left(\hat{b}_p^\dagger \hat{b}_p + \hat{d}_p^\dagger \hat{d}_p\right) dp - \int_{-\infty}^{+\infty} E_p dp + \xi_R = \int_{-\infty}^{+\infty} E_p \left(\hat{b}_p^\dagger \hat{b}_p + \hat{d}_p^\dagger \hat{d}_p\right) dp \tag{19}$$

where the last step was achieved by defining the renormalization constant so that $\xi_R - \int_{-\infty}^{+\infty} E_p dp = 0$. It is evident that,

$$\hat{H}_0 \left|0\right\rangle = \xi\left(\left|0\right\rangle\right)\left|0\right\rangle = 0 \tag{20}$$

where $\xi\left(\left|0\right\rangle\right) = 0$ is the energy of the vacuum state. Additional eigenstates $\left|n\right\rangle$ are formed by acting on the vacuum state $\left|0\right\rangle$ with the various combinations of the creation operators $b_p^\dagger$ and $d_p^\dagger$. The effect of doing this is to create states with positive energy with respect to the vacuum state. The set of eigenstates $\left|n\right\rangle$ (which includes the vacuum state $\left|0\right\rangle$) form an orthonormal basis that satisfies the following relationships,

$$\hat{H}_0 \left|n\right\rangle = \xi\left(\left|n\right\rangle\right)\left|n\right\rangle \text{ where } \xi\left(\left|n\right\rangle\right) > \xi\left(\left|0\right\rangle\right) \text{ for } \left|n\right\rangle \neq \left|0\right\rangle \tag{21}$$

and

$$\left\langle n|m\right\rangle = \delta_{mn}; \quad \sum_{\left|n\right\rangle} \left|n\right\rangle\left\langle n\right| = 1 \tag{22}$$

### 3. An anomaly.

The current and charge operators in 1-1dimensional space-time are $\hat{\rho}(z) = \hat{\psi}^\dagger(z)\hat{\psi}(z)$ and $\hat{J}(z) = \hat{\psi}^\dagger(z)\sigma_x\hat{\psi}(z)$. Using this along with (9) it can be shown that,

$$\left[\hat{H}_0, \hat{\rho}(z)\right] = i\frac{\partial \hat{J}(z)}{\partial z} \tag{23}$$

This is just the continuity equation in 1-1D space-time.

Now consider the quantities $I_1$ and $I_2$ defined by,

$$I_1 = \left\langle 0 \right| \left[ \hat{F}, \left[ \hat{H}_0, \hat{F} \right] \right] \left| 0 \right\rangle \text{ and } I_2 = \left\langle 0 \right| \left[ \hat{F}, \left[ \hat{H}_0, \hat{F} \right] \right] \left| 0 \right\rangle \tag{24}$$

where,

$$\hat{F} = \int \hat{\rho}(z) f(z) dz \tag{25}$$

and where $f(z)$ is an arbitrary real-valued function. It is obvious that $I_1$ and $I_2$ should be equal. However we will use different methods to determine each quantity. We will find that when we do this the two quantities are not equal. Therefore we have an anomaly.

In order to calculate $I_1$ we will use the anti-commutation relationships (9). First use (23) to obtain,

$$\left[ \hat{H}_0, \hat{F} \right] = i \int \frac{\partial \hat{J}(z)}{\partial z} f(z) dz = -i \int \frac{\partial f(z)}{\partial z} \hat{J}(z) dz \tag{26}$$

Use this in (24) to obtain,

$$I_1 = -i \left\langle 0 \right| \iint f(z) \frac{\partial f(z')}{\partial z'} \left[ \hat{\rho}(z), \hat{J}(z') \right] dz' dz \left| 0 \right\rangle \tag{27}$$

Use (10) (which is derived from (9)) to obtain,

$$\left[ \hat{\rho}(z), \hat{J}(z') \right] = \left( \hat{\psi}^\dagger(z) \sigma_x \hat{\psi}(z') - \hat{\psi}^\dagger(z') \sigma_x \hat{\psi}(z) \right) \delta(z - z') \tag{28}$$

Use this in (27) to obtain,

$$I_1 = -i \left\langle 0 \right| \iint f(z) \frac{\partial f(z')}{\partial z'} \left( \hat{\psi}^\dagger(z) \sigma_x \hat{\psi}(z') - \hat{\psi}^\dagger(z') \sigma_x \hat{\psi}(z) \right) \delta(z - z') dz' dz \left| 0 \right\rangle \tag{29}$$

Perform the integration with respect to $z'$ to obtain,

$$I_1 = -i \left\langle 0 \right| \int f(z) \frac{\partial f(z)}{\partial z} \left( \hat{\psi}^\dagger(z) \sigma_x \hat{\psi}(z) - \hat{\psi}^\dagger(z) \sigma_x \hat{\psi}(z) \right) dz \left| 0 \right\rangle = 0 \tag{30}$$

To obtain the above result we have used the definition of the delta function $\int g(z') \delta(z - z') dz' = g(z)$.

Next we will evaluate $I_2$ using a different method. We will find that we obtain a different result. Refer to (24) and expand the commutation and use $\hat{H}_0 \left| 0 \right\rangle = \left\langle 0 \right| \hat{H}_0 = 0$ to obtain,

$$I_2 = 2 \left\langle 0 \right| \hat{F} \hat{H}_0 \hat{F} \left| 0 \right\rangle \tag{31}$$

Next use (21) and (22) to obtain,

$$I_2 = 2\sum_{|n\rangle}\sum_{|m\rangle}\langle 0|\hat{F}|n\rangle\langle n|\hat{H}_0|m\rangle\langle m|\hat{F}|0\rangle = 2\sum_{|m\rangle}\left|\langle 0|\hat{F}|m\rangle\right|^2 \xi\left(|m\rangle\right) > 0 \qquad (32)$$

The reason the above quantity is greater than zero is because $\left|\langle 0|\hat{F}|m\rangle\right|^2$ is obviously non-negative and is, in general, non-zero and $\xi\left(|m\rangle\right) > 0$ if $|m\rangle$ is not equal to the vacuum state. This result is obviously not consistent with (30). We have obtained $I_1 \neq I_2$ even though $I_1$ and $I_2$ were originally defined to be identical quantities. Therefore we have an anomaly due to the fact that two different ways of working the problem give different results.

## 4. Point splitting the current operator.

Now how can this situation be resolved? One way is to try to figure out how to make $I_2$ equal to zero. This can only happen if states exist with less energy than the vacuum state. A way to define the vacuum state so that it is no longer the lower bound to the energy has been suggested by the author (D. Solomon [5]). This may seem somewhat a radical concept however it has been recently shown that in Dirac Hole theory there exist states with less energy than the vacuum state [6][7]. However, the more traditional approach is to maintain the notion that the vacuum is the minimum energy state. This, then, means that we preserve the result $I_2 > 0$ and try to modify the theory so that $I_1$ is consistent with this.

One method to achieve this result is to modify the definition of the current operator by using point splitting. Define the point split current operator as follows,

$$\hat{J}(z;\varepsilon) = \frac{1}{2}\sum_{\gamma=\pm 1}\hat{\psi}^\dagger(z+\gamma\varepsilon)\sigma_x\hat{\psi}(z) \qquad (33)$$

where $\varepsilon \to 0$ but is never set equal to zero. Use this in (27) to obtain,

$$I_1(\varepsilon) = -i\langle 0|\iint f(z)\frac{\partial f(z')}{\partial z'}\left[\hat{\rho}(z),\hat{J}(z';\varepsilon)\right]dz'dz|0\rangle \qquad (34)$$

Refer to (10) to obtain,

$$\left[\hat{\rho}(z),\hat{J}(z';\varepsilon)\right] = \frac{1}{2}\sum_{\gamma=\pm 1}\begin{Bmatrix}\hat{\psi}^\dagger(z)\sigma_x\hat{\psi}(z')\delta\left(z-(z'+\gamma\varepsilon)\right)\\-\hat{\psi}^\dagger(z'+\gamma\varepsilon)\sigma_x\hat{\psi}(z)\delta(z-z')\end{Bmatrix} \qquad (35)$$

Use this in (34) to yield,

$$I_1(\varepsilon) = \frac{-i}{2} \sum_{\gamma=\pm 1} \iint f(z) \frac{\partial f(z')}{\partial z'} \langle 0 | \begin{pmatrix} \hat{\psi}^\dagger(z) \sigma_x \hat{\psi}(z') \delta(z - (z' + \gamma\varepsilon)) \\ -\hat{\psi}^\dagger(z' + \gamma\varepsilon) \sigma_x \hat{\psi}(z) \delta(z - z') \end{pmatrix} | 0 \rangle \, dz' dz \quad (36)$$

Integrate to obtain,

$$I_1(\varepsilon) = \frac{-i}{2} \sum_{\gamma=\pm 1} \left\{ \begin{array}{l} \int f(z) \dfrac{\partial f(z - \gamma\varepsilon)}{\partial z} \langle 0 | \hat{\psi}^\dagger(z) \sigma_x \hat{\psi}(z - \gamma\varepsilon) | 0 \rangle \, dz \\ -\int f(z) \dfrac{\partial f(z)}{\partial z} \langle 0 | \hat{\psi}^\dagger(z + \gamma\varepsilon) \sigma_x \hat{\psi}(z) | 0 \rangle \, dz \end{array} \right\} \quad (37)$$

Make the transformation $z \to z + \gamma\varepsilon$ in the first integral in the above equation to obtain,

$$I_1(\varepsilon) = \frac{-i}{2} \sum_{\gamma=\pm 1} \int \left( f(z + \gamma\varepsilon) - f(z) \right) \frac{\partial f(z)}{\partial z} \langle 0 | \hat{\psi}^\dagger(z + \gamma\varepsilon) \sigma_x \hat{\psi}(z) | 0 \rangle \, dz \quad (38)$$

Use (11) and (14) to yield,

$$\langle 0 | \hat{\psi}^\dagger(z + \gamma\varepsilon) \sigma_x \hat{\psi}(z) | 0 \rangle = \int_{-\infty}^{+\infty} \left( \frac{-p}{E_p} \right) e^{-ip\gamma\varepsilon} \, dp = 4i \int_0^{+\infty} \left( \frac{p}{E_p} \right) \sin(p\gamma\varepsilon) \, dp \quad (39)$$

Use this result in (38) to obtain,

$$I_1(\varepsilon) = 2 \sum_{\gamma=\pm 1} \int \left( f(z + \gamma\varepsilon) - f(z) \right) \frac{\partial f(z)}{\partial z} \, dz \int_0^{+\infty} \left( \frac{p}{E_p} \right) \sin(p\gamma\varepsilon) \, dp \quad (40)$$

Now we need to evaluate the above expression. In the limit that $\varepsilon \to 0$ the quantity $\sin(p\gamma\varepsilon)$ approaches zero unless $p \to \infty$. Therefore we can replace $E_p$ with $|p|$ to obtain,

$$\int_0^{+\infty} \frac{p}{E_p} \sin(\gamma p\varepsilon) \, dp \underset{\varepsilon \to 0}{=} \int_0^{+\infty} \sin(\gamma p\varepsilon) \, dp = P\left( \frac{1}{\gamma\varepsilon} \right) = \begin{cases} 1/\gamma\varepsilon & \text{if } \varepsilon \neq 0 \\ 0 & \text{if } \varepsilon = 0 \end{cases} \quad (41)$$

Use this in (40) and let $\varepsilon \to 0$ to obtain,

$$I_1(\varepsilon) \underset{\varepsilon \to 0}{=} 2 \sum_{\gamma=\pm 1} \int \left( \frac{f(z + \gamma\varepsilon) - f(z)}{\gamma\varepsilon} \right) \frac{\partial f(z)}{\partial z} \, dz \quad (42)$$

Next use $\left( \dfrac{f(z + \gamma\varepsilon) - f(z)}{\gamma\varepsilon} \right) \underset{\varepsilon \to 0}{=} \dfrac{df(z)}{dz}$ in the above to obtain,

$$I_1(\varepsilon) \underset{\varepsilon \to 0}{=} 2 \sum_{\gamma=\pm 1} \int \left( \frac{\partial f(z)}{\partial z} \right)^2 \, dz > 0 \quad (43)$$

This equation is consistent with (32). Therefore the use of point splitting has removed the anomaly.

## 5. Point Splitting the Hamiltonian

As stated in the introduction there is a problem with defining the current operator per equation (33). This is because equation (23) must still hold. Equation (23) is the continuity equation which is an expression of local charge conservation. In order that the continuity equation is consistent with (33) the quantity $\varepsilon$ must appear on the left side of (23). D. Boulware[4] argues that the Hamiltonian should be point split and that the charge operator should remain unchanged.

Following this prescription we define the point split Hamiltonian according to,

$$\hat{H}_{0,\varepsilon} = \frac{1}{2}\sum_{\gamma=\pm 1}\int \hat{\psi}^{\dagger}(z+\gamma\varepsilon)H_0\hat{\psi}(z)dz + \xi_{R,\varepsilon} \tag{44}$$

where we take the limit $\varepsilon \to 0$ (but do not allow $\varepsilon$ to equal 0). The current operator is defined by the relationship,

$$\left[\hat{H}_{0,\varepsilon},\hat{\rho}(z)\right] = i\frac{\partial \hat{J}(z;\varepsilon)}{\partial z} \tag{45}$$

Use the above relationships along with (16) to obtain,

$$\left[\hat{H}_{0,\varepsilon},\hat{\rho}(z)\right] = \frac{1}{2}\sum_{\gamma=\pm 1}\left(\begin{array}{c} i\left(\dfrac{\partial \hat{\psi}^{\dagger}(z+\gamma\varepsilon)}{\partial z}\sigma_x\hat{\psi}(z) + \hat{\psi}^{\dagger}(z)\sigma_x\dfrac{\partial \hat{\psi}(z+\gamma\varepsilon)}{\partial z}\right) \\ +m\left(\hat{\psi}^{\dagger}(z+\gamma\varepsilon)\sigma_z\hat{\psi}(z) - \hat{\psi}^{\dagger}(z)\sigma_z\hat{\psi}(z+\gamma\varepsilon)\right) \end{array}\right) \tag{46}$$

From this result and (45) we can show that,

$$\hat{J}(z;\varepsilon) = \frac{1}{2}\sum_{\gamma=\pm 1}\left\{\hat{\psi}^{\dagger}(z+\gamma\varepsilon)\sigma_x\hat{\psi}(z) + i\int_{z-\gamma\varepsilon}^{z}\hat{\psi}^{\dagger}(z'+\gamma\varepsilon)\left(i\sigma_x\frac{\partial}{\partial z'} - m\sigma_z\right)\hat{\psi}(z')dz'\right\} \tag{47}$$

Note that this current operator is somewhat more complicated then (33).

## 6. Point splitting and the vacuum state.

Next we will examine the effect of point splitting on the definition of the vacuum state. Use (11) in (44) to obtain,

$$\hat{H}_{0,\varepsilon} = \frac{1}{2}\sum_{\gamma=\pm 1}\int_{-\infty}^{+\infty}E_p\left(\hat{b}_p^{\dagger}\hat{b}_p - \hat{d}_p\hat{d}_p^{\dagger}\right)e^{-i\gamma p\varepsilon}dp + \xi_{R,\varepsilon} = \int_{-\infty}^{+\infty}E_p\left(\hat{b}_p^{\dagger}\hat{b}_p - \hat{d}_p\hat{d}_p^{\dagger}\right)\cos(p\varepsilon)dp + \xi_{R,\varepsilon} \tag{48}$$

Next use the anti-commutation relationships (12) to obtain

$$\hat{H}_{0,\varepsilon} = \int_{-\infty}^{+\infty} E_p \left( \hat{b}_p^\dagger \hat{b}_p + \left( \hat{d}_p^\dagger \hat{d}_p - 1 \right) \right) \cos\left( p\varepsilon \right) dp + \xi_{R,\varepsilon} = \int_{-\infty}^{+\infty} E_p \left( \hat{b}_p^\dagger \hat{b}_p + \hat{d}_p^\dagger \hat{d}_p \right) \cos\left( p\varepsilon \right) dp \quad (49)$$

where we define $\xi_{R,\varepsilon}$ such that $\xi_{R,\varepsilon} = \int_{-\infty}^{+\infty} E_p \cos\left( p\varepsilon \right) dp$. Using (49) it is evident that $\hat{H}_{0,\varepsilon} \left| 0 \right\rangle = 0$. Therefore the energy of the vacuum state $\left| 0 \right\rangle$ is zero which is consistent with the discussion in Section 2 (see Eq. (20)).

However there is a potential problem. It is easy to show that there are states with less energy than the state $\left| 0 \right\rangle$. For example, consider the energy of the state $\hat{b}_q^\dagger \left| 0 \right\rangle$ or $\hat{d}_q^\dagger \left| 0 \right\rangle$. We have,

$$\left\langle 0 \right| \hat{b}_q \hat{H}_{0,\varepsilon} \hat{b}_q^\dagger \left| 0 \right\rangle = \left\langle 0 \right| \hat{d}_q \hat{H}_{0,\varepsilon} \hat{d}_q^\dagger \left| 0 \right\rangle = E_q \cos q\varepsilon \qquad (50)$$

Note that if $\cos q\varepsilon < 0$ the energy of this state will be negative. Even though $\varepsilon$ is arbitrarily small the momentum $q$ can be arbitrarily large. Therefore for any $\varepsilon$ there always exists a momentum $q$ such that the quantity $q\varepsilon$ is finite and $\cos\left( q\varepsilon \right)$ is negative. Therefore the vacuum state $\left| 0 \right\rangle$ is not the state of lowest energy.

At this point there is a problem because in point splitting the Hamiltonian we have lost the fact that the vacuum state $\left| 0 \right\rangle$ is the minimum energy state. However it is possible to regain the concept of a minimum energy state by redefining the vacuum. Define the set $S_\varepsilon$ as the set of all $q$ such that $\cos q\varepsilon < 0$. Define the state $\left| 0, \varepsilon \right\rangle$ according to,

$$\left| 0, \varepsilon \right\rangle = \prod_{q \in S_\varepsilon} \hat{d}_q^\dagger \hat{b}_q^\dagger \left| 0 \right\rangle \qquad (51)$$

That is, all negative energy states are occupied by an electron or positron. This is similar to Dirac's hole theory where the negative energy states are all occupied in order to produce the vacuum state. This is the main result of this paper. If we want to formulate a consistent theory of point splitting we must reformulate the definition of the vacuum state per Eq. (51). If this is not done then there will exist states with less energy than the vacuum state.

**7.  Conclusion**

We have a examined some problems associated with trying to achieve a consistent theory using point splitting.  We have followed the suggestion of D.G. Boulware [4] that it is necessary to apply point splitting to the Hamiltonian operator and define the current operator according to (45).  When this is done we achieve consistency in the theory in that $I_1$ and $I_2$ are both greater than zero.  However when we examine the effect of point splitting on the Hamiltonian it is seen that there are states with less energy than the vacuum state $\left|0\right\rangle$.  The vacuum state must be redefined as specified by Eq. (51) in order to produce a minimum energy state.